\documentstyle[11pt,newpasp,twoside,epsf]{article} 
\def\edcomment#1{\iffalse\marginpar{\raggedright\sl#1\/}\else\relax\fi} 
\reversemarginpar 

\begin{document} 
\title{Comparing the old Stellar Population in Globulars 
and Dwarf Galaxies: The Cases of Phoenix and Leo A}

\author{Ulrich Hopp} 
\affil{Universit\"ats-Sternwarte M\"unchen, Scheiner Str. 1, D 81679
Munich, Germany} 
\author{Laura Greggio} 
\affil{INAF, Osservatorio Astronomico di Padova, Vicolo dell'Osservatorio 5,
Padova, Italy} 

\begin{abstract} 
Due to their star formation history (SFH), the stellar population in Dwarf
Galaxies (DG) is likely to have a metallicity spread which is best traced
by the morphology of the Red Giant Branch (RGB). We probe here a purely
empirical approach aimed at estimating average metallicity ($Z$) and $Z$
spread by comparing the Color-Magnitude Diagrams (CMD) of galactic
Globular Clusters (GCs) with those of two DGs:  Leo A (HST data) and 
Phoe (VLT Fors2 data). 
\end{abstract} 


The older ($>$~1~Gyr) stellar population of nearby galaxies 
holds a very important information of the SFH of the local universe.
This is often derived by comparing the observed CMDs
with theoretical simulations based on isochrones (e.g. Tosi, 2000).
However, theoretical isochrones have some difficulty in reproducing the
overall appearence of the RGBs of globular clusters and their systematics
with metallicity. As an alternative method we propose to use observed
CMDs of GCs as {\em empirical} Simple Stellar Populations (SSP), i.e.
assembly of coeval stars all with the same metallicity $Z$.

As test cases, we select the HST observations by Schulte-Ladbeck et al. 
(2002) of the outer envelope of Leo A and our own VLT FORS2 observations of
the Phoenix dwarf galaxy. Both galaxies have undergone an extended period of 
SF, as evidenced from the presence of an extended horizontal branch as well
as younger stars (in their centers). Thus, both systems are an example of 
Composite Stellar Populations (CSP).
The globular cluster comparison set comes from
Rosenberg et al. (1999), and encompasses the metallicity range 
-2.2 $<$ [Fe/H] $<$ -0.7. 
Since this set is limited to old ages, we only consider the
outer parts of the galaxies, dominated by the old component of the CSP.

In Fig.~1 (left panel) we superimpose the CMD of Phoenix  to the CMDs of
the GCs, with [Fe/H]= -1.6. The morphology of the galaxy's RGB and HB are 
very well matched. The results of Kolmogorov-Smirnov (KS) tests applied to 
the stellar distribution of the two DGs against the GCs of various
metallicities are shown in the right panel of Fig.~1. 
The comparisons indicate an average metallicity (spread) of [Fe/H] = -1.8 
(0.8 dex) for Leo A and -1.6 (0.6 dex) for Phoenix.
  
As a further step one can compare the galaxies' data to a 
CSP constructed with a set of GCs, taken in a suitable
combination. For an SSP, the number of stars populating a given Post Main 
Sequence (PMS) phase is proportional to the total mass of the SSP 
($M_{\rm S}$) through a factor $\delta n_{\rm S}$, which 
is a robust prediction from stellar evolution models (Greggio 2002, 
astro-ph/0111241).
Thus, if $\Delta N$ is the total nmber of stars of the empirical CSP
in a selected PMS phase (e.g. the RGB), each cluster contributes 
\begin{equation}
\frac{\Delta N_{\rm C}}{ \Delta N} = \frac{f_{\rm S}\times \delta n_{\rm S}}
{\Sigma_{\rm S}\, f_{\rm S}\times \delta n_{\rm S}}
\end{equation}
RGB stars, where $f_{\rm S}$ is the contribution
by mass of cluster $S$ (to be adopted), and the sum is performed on all 
the clusters composing the empirical CSP. This produces the RGB of the CSP.
The other evolutionary phases can be obtained by simply scaling to 
$\delta n_{\rm S}$ the original proportions in the GC CMD.

For old stellar populations $\delta n_{\rm S}$ mostly depends on
$Z$ and on the Initial Mass Function slope ($\alpha$). 
As an example,  for RGB stars brighter than $M_{\rm I}$=-1.5, 
$\delta n_{\rm S}$ decreases by 
a factor of 0.6 when $[Fe/H]$ increases from -2.4 to 0. At solar
metallcity, $\delta n_{\rm S}$ increases by a factor of 1.2 when $\alpha$ 
varies from -2.35 to -1.3. 

\begin{figure}
\plottwo{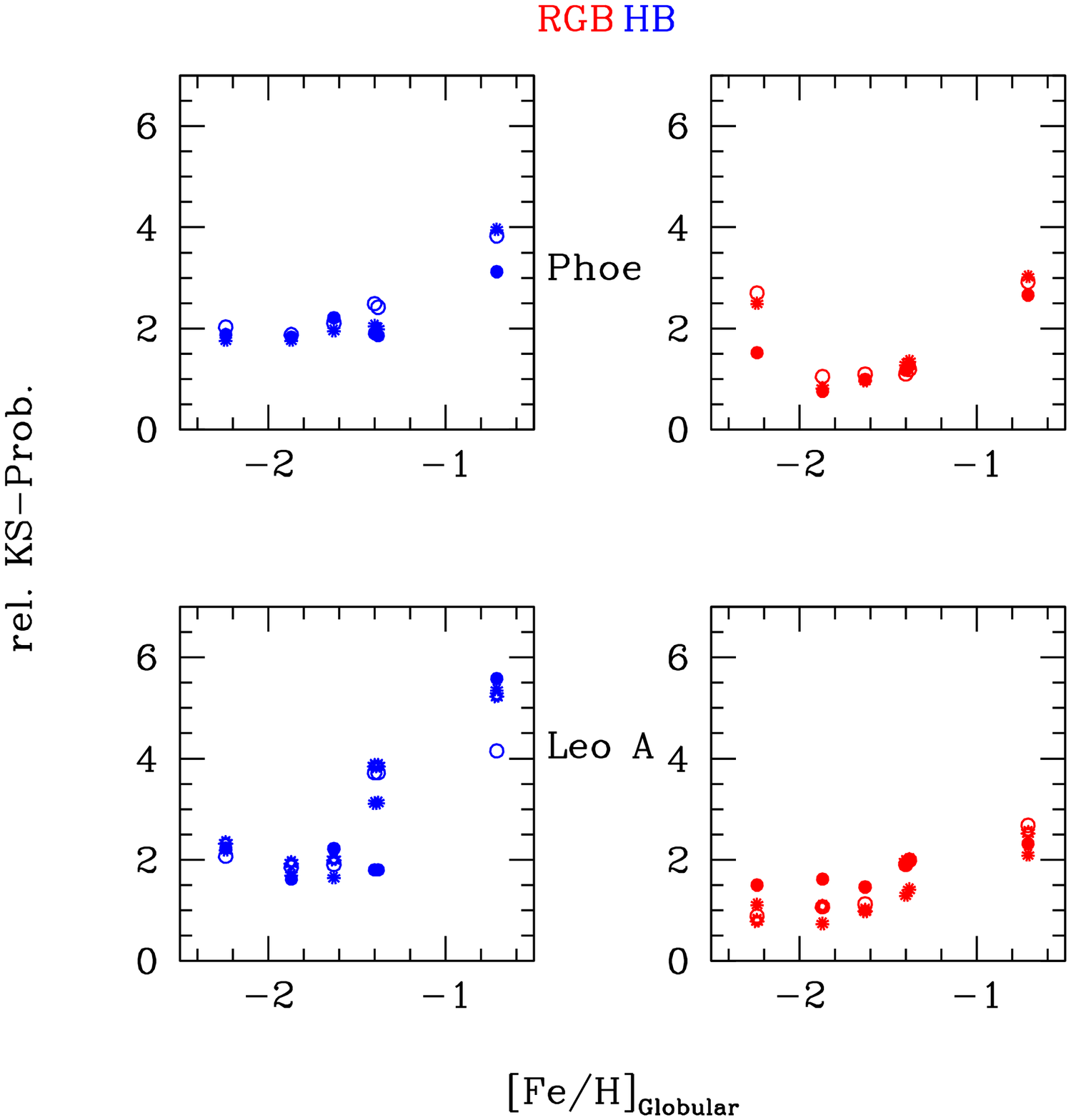}{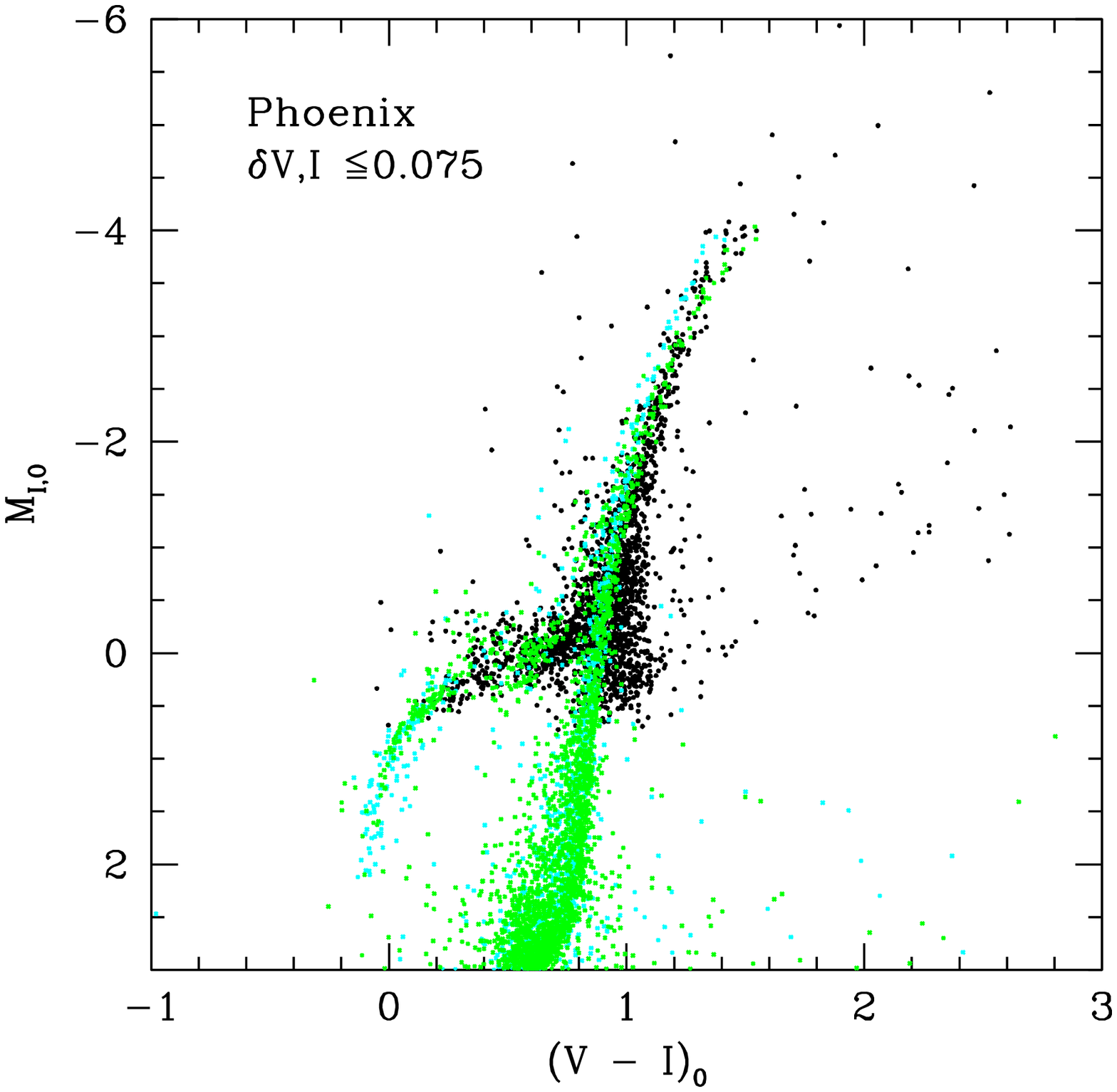}
\caption{ Left Panel: comparison of selected GCs CMDs with the data for 
Phoenix (outer part only, no recent SF).
Right Panel: results of KS tests applied to Phoenix (upper) and Leo A (lower).
The left frames show the result for the color distribution of HB stars
with 0.2 $< V-I <$ 0.8. The right frames show the results for the 
magnitude distribution of RGB stars with 0.9 $< V-I <$ 1.6.} 
\end{figure}


\begin{references}

\verb"Rosenberg, A. et al. 2000, A\&A, 144, 5"

\verb"Schulte-Ladbeck,R.E. et al. 2002, AJ, 124, 896"

\verb"Tosi, M. 2000 in: The Magellanic Clouds and other dwarf galaxies,"\\
\verb"eds. K. S. De Boer, R.-J. Dettmar, U. Klein, Shaker Verlag, p. 67"

\end{references}
\end{document}